\documentclass[a4paper,11pt]{article}
\usepackage{pos}
\usepackage{slashed}

\title{First dynamical simulations with minimally doubled fermions}

\author*[a]{Reka A. Vig}
\author[a]{Szabolcs Borsanyi}
\author[a,b]{Zoltan Fodor}
\author[b]{Daniel Godzieba}
\author[b]{Paolo Parotto}
\author[a]{Chik Him Wong}

\affiliation[a]{Physics department, University of Wuppertal,\\
  Gaussstrasse 20., Wuppertal, Germany}

\affiliation[b]{Physics department, Pennsylvania State University,\\
 0104 Davey Laboratory University Park, Pennsylvania, USA}

\emailAdd{tajhajlito@gmail.com}
\emailAdd{borsanyi@uni-wuppertal.de}
\emailAdd{zxf5098@psu.edu}
\emailAdd{dag5611@psu.edu}
\emailAdd{paolo.parotto@gmail.com}
\emailAdd{cwong@uni-wuppertal.de}

\abstract{For thermodynamics studies it is desirable to simulate two degenerate flavors and retain at least a remnant of the chiral
symmetry. Staggered fermions can achieve this at the cost of rooting the determinant. Rooting can be
avoided using minimally doubled fermions. This discretization describes two degenerate quark
flavors while explicitly breaking hyper-cubic symmetry, thus, requiring additional counter-terms.
We use one particular formulation of minimally doubled fermions called the Kirsten-Wilczek
action and mitigate lattice artifacts by improving the spatial derivatives in the Dirac operator. In
this pilot study we determine the counter-terms non-perturbatively to facilitate proper dynamical
simulations.}

\FullConference{The 40th International Symposium on Lattice Field Theory (Lattice 2023)\\
July 31st - August 4th, 2023\\
Fermi National Accelerator Laboratory\\}

\begin{document}
\maketitle

\section{Introduction}
Choosing the right action is essential in dynamical lattice simulations. To study the chiral properties of a thermodynamical system we have to utilize an action which at least preserve a remnant chiral symmetry. Chiral actions are numerically expensive among which Staggered fermions is the cheapest, hence it is the most widely used type of discretization. Working with less than four degenerate quark flavors is done by rooting the fermion determinant. Rooting becomes problematic at finite real chemical potential ($\mu$) since we have to take the square root of a complex determinant which introduces a sign ambiguity. This becomes severe at chemical potentials near the complex zero of the determinant. This ambiguity can cause unwanted effects in thermodynamic observables (for details see Ref. \cite{Borsanyi:2023tdp}). 
Overlap fermions is a doubler free discretization that satisfies the Ginsparg-Wilson relation hence retain chiral symmetry on the lattice. However, it is numerically costly and at a $\mu \neq 0$ it is complicated to determine the sign function required in the formula of the operator.
The above reasons motivated us to explore the possibilities of minimally doubled fermions \cite{Karsten,Wilczek}. This type of discretization realizes two degenerate quark flavors in the coninuum,  besides it is ultra-local and retains a remnant chiral symmetry. Its shortcoming is that it explicitly breaks the hyper-cubic symmetry.Thus it is necessary to introduce counter-terms \cite{Capitani:2010ht} to the bare action to obtain a properly renormalized theory. The form of the counter-terms can be determined using perturbation theory and their coefficients must be tuned non-perturbatively to restore hyper-cubic symmetry on the lattice.
In this pilot study we tune the coefficients of the three counter-terms of a particular type of minimally doubled action called the Karsten-Wilczek (KW) action. 
\section{Karsten-Wilczek fermions} \label{KW}
The simplest kind of minimally doubled fermions was first proposed by Karsten \cite{Karsten} and Wilczek \cite{Wilczek}. Karsten's solution to remove fourteen of the fifteen spurious doublers was to add a term to the naive fermion action that anticommutes with $\gamma_5$ hence does not violate chiral symmetry. The KW term only exists in three of the four space-time directions, thus, only one doubler remains which lies in the fourth space-time direction in the Brillouin zone. This means the KW term breaks the hyper-cubic symmetry to cubic symmetry of the subspace of the three doubler free directions. Wilczek generalized the action by introducing the so called Wilczek parameter ($\zeta$). The tree level Karsten-Wilczek action reads
\begin{equation}
S_F^{KW} = S_F^N +  \sum_x \sum_{j\neq \alpha}\bar{\psi}(x) \frac{i \zeta}{2} \gamma^\alpha \left( 2 \psi(x) - U_j(x) \psi(x+\hat{j}) - U_j^\dagger(x-\hat j) \psi(x-\hat{j})  \right) \rm{,}
\end{equation}
where $S_F^N$ is the naive fermion action
\begin{equation}
S_F^N=\sum_x \sum_{\mu=0}^3 
\bar\psi(x) \gamma_\mu \frac12 \left[U_\mu(x)
\psi(x+\hat\mu) - U^+_\mu(x-\hat\mu)\psi(x-\hat\mu)
\right] + m \sum_x \bar\psi(x)\psi(x)
\end{equation} 
where $U_\mu(x)$ are the gauge links in direction $\mu$ at lattice site $x$ and $\alpha$ can be any of the four directions. The KW term is similar to the Wilson term but an important difference is that it is multiplied by $\gamma_\alpha$. 
For a proper renormalization of the KW action we need to introduce two fermionic counter-terms (of dimensions 3 and 4 denoted by $S^{3f}$ and $S^{4f}$) and one gluonic 
counter-term (of dimension 4 denoted by $S^{4g}$)
\begin{equation}
\begin{aligned}
S^{3f} &= c \sum_x \bar{\psi}(x) i \gamma^\alpha \psi(x)\rm{,} \quad S^{4g} = d_G \sum_x \sum_{\mu\ne\alpha} \operatorname{Re} \operatorname{Tr} \left( 1 - \mathcal{P}_{\mu\alpha}(x)\right)\rm{,} \\
S^{4f} &= d \sum_x \bar{\psi}(x) \frac{1}{2} \gamma^\alpha \left( U_\alpha(x)\psi(x+\hat{\alpha}) - U_\alpha^\dagger(x-\hat{\alpha})\psi(x-\hat{\alpha})) \right)\rm{,} 
\end{aligned}
\label{eq:ct}
\end{equation}
where $\mathcal{P}_{\mu\alpha}(x)$ are the plaquettes in direction $\alpha$ at lattice site $x$ and $c, d$ and $d_G$ are the coefficients to be tuned. 
We chose $\alpha$ to be the time direction (denoted by $0$) which we will use from now on.

\begin{figure}
 \centering 
\includegraphics[width=.495\textwidth]{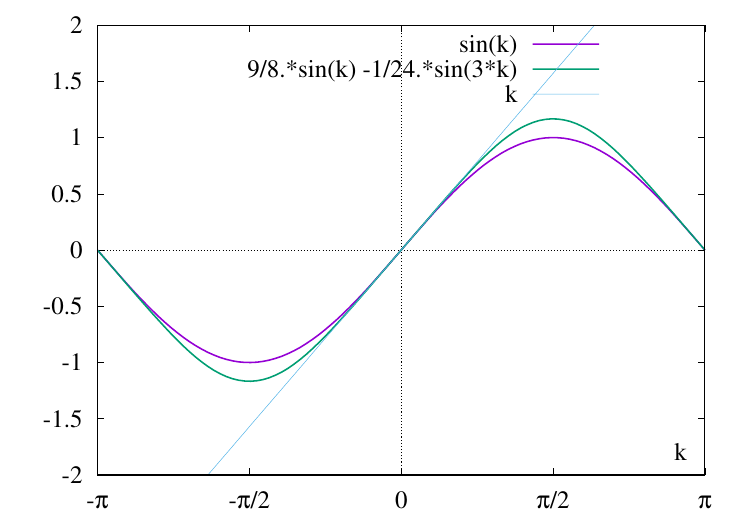} \includegraphics[width=.495\textwidth]{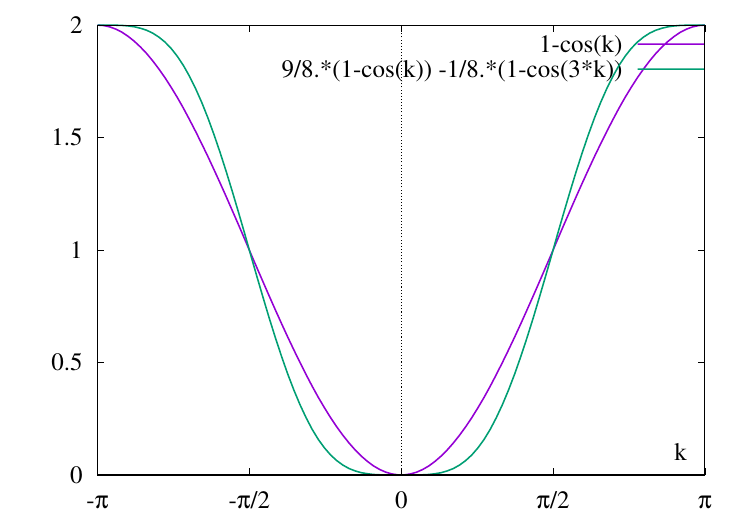}
           \caption{The Nabla (left hand side) and Laplacian (right hand side) terms of the momentum space Dirac operator in the unimproved (purple line), improved (green line) and continuum (blue line) case as a function of the momentum.\label{fig:naik_momentum} }
\end{figure}

To mitigate the effects of finite lattice spacing we also introduced additional terms to improve the action. A simple and perhaps cheap choice is the so-called Naik improvement \cite{Naik:1986bn}. To demonstrate it we start from the momentum space expression of the KW Dirac operator 
\begin{equation}
    D_{\rm KW}(k) = \frac{i}{a} \left[
\sum_{\mu=0}^3 \gamma_\mu \xi_\mu \sin\,ak_\mu +\zeta\gamma_0 \sum_{j=1}^3(1-\cos\, ak_j) \right]
\end{equation}
where the $\sin\,a k_j$ corresponds to a Nabla term and the $(1-\cos\,ak_j)$ is a Laplacian term in the direction $j$, similarly to the Wilson operator. 
We applied the Naik improvement in the three spatial directions. Since the action is inherently anisotropic we can account for the lack of improvement in time direction by using a smaller lattice spacing in time.   
The Naik improved Nabla term is the following
\begin{equation} \label{nabla}
\nabla^{\rm NAIK}_j \Psi =  \frac{i}{a}\left[s_1 \sin\,a k_j + s_3\sin\,3a k_j\right]\Psi(k) = i\left[ k_j + 0 k_j^3 +\mathcal{O}(k^5)\right].
\end{equation}
With the choice  $s_1=9/8$, $s_3=-1/24$ the criterion in Eq. (\ref{nabla}) is satisfied .
With the improvement of the Laplacian 
\begin{equation}
\Delta^{\rm NAIK}_j \Psi = \frac{i}{a}\left[ c_1 (1-\cos\,a k_j) + c_3 (1-\cos\,3a k_j) \right]\Psi(k)
\end{equation}
we want to suppress the momentum dependence of the cosine term and simultaneously lift
the momenta in the doubler's part of the Brillouin zone. This is realized with the choice $c_1=9/8$, $c_3=-1/8$, illustrated on the right of Fig.~\ref{fig:naik_momentum}.
With these choices of the coefficients the half vector trick becomes impossible, thus, one has to decide whether it is worth to sacrifice the exact improvement for faster computations. Our simulations were done by using the half vector trick. We set the coefficients considering the following. Let us take a look at the Naik improved KW Dirac operator in coordinate space 
\begin{align}\label{improvedD}
D \psi&[s] =
\sum_\mu \xi_\mu\left[
c_\mu(s)~\Gamma_{\rm (1)}^\mu~U_\mu(s)~\psi[s+\mu] \right.\\
&
+c_\mu(s)c_\mu(s+\mu)c_\mu(s+2\mu)~\Gamma_{\rm (3)}^\mu~
U_\mu(s)U_\mu(s+\mu)U_\mu(s+2\mu)
~\psi[s+3\mu] \label{naik_f}\\
&
- c^{-1}_\mu(s-\mu)~ \Gamma_{\rm(1)}^{\mu\dagger} U^\dagger_\mu(s-\mu)\psi[s-\mu]\\
&
\left.
- c^{-1}_\mu(s-\mu)c^{-1}_\mu(s-2\mu)c^{-1}_\mu(s-3\mu)~ \Gamma_{\rm(3)}^{\mu\dagger} 
U^\dagger_\mu(s-\mu)U^\dagger_\mu(s-2\mu)U^\dagger_\mu(s-3\mu)
\psi[s-3\mu]\right] \label{naik_b}
\\
&+(2m + 2i(3\zeta+c)\gamma^0)~\psi[s]
\end{align}
where the \ref{naik_f} and \ref{naik_b} terms correspond to the 3D Naik improvement and
\begin{align}\label{Gamma}
\Gamma_{\rm (1)}^\mu &= \bar s_1 \gamma^\mu  - i \bar c_1 \gamma^0,\quad
\Gamma_{\rm (3)}^\mu = \bar s_3 \gamma^\mu  - i \bar c_3 \gamma^0 \quad (\mu\ne 0);\quad\Gamma_{\rm (1)}^0 = (1+d)\gamma^0, \quad
\Gamma_{\rm (3)}^0 = 0 
\end{align}
where $
1 = \bar s_1 + 3\bar s_3,~
\zeta =\bar c_1 + \bar c_3 ~
\mathrm{with}~\bar c_1 = \zeta c_1, ~
\bar c_3=\zeta c_3 ~ \mathrm{and}~ \bar s_1 =s_1, ~
\bar s_3=s_3.
$
One can see from the structure of $\Gamma_{(1),(3)}$ in Eq. (\ref{Gamma}) that we have to require $\bar s_1=\bar c_1$ and $\bar s_3=\bar c_3$ to make the half vector trick work. We want to keep the exact Laplacian improvement, thus we changed the previously determined value of $s_1$ and $s_3$. We set the coefficients to be
$\bar s_1=\bar c_1=1.5,~ \bar s_3=\bar c_3=-1/6$ and $\zeta =4/3.$

The properties of the KW Dirac operator make it possible to realize simulations using the Hybrid Monte Carlo algorithm (HMC) without rooting.
With the notation $D^{KW} = m + \slashed{D}^{KW}$ these are
\begin{align}
&\slashed{D}^{KW\dag} = \gamma_5\slashed{D}^{KW}\gamma_5;\quad
D^{KW\dag} \gamma_5 = \gamma_5 D^{KW} \quad \gamma_5\mathrm{-hermiticity,}\\
&\gamma_5\slashed{D}^{KW}\gamma_5 = -\slashed{D}^{KW}
\rightarrow D^{KW\dag} = m - \slashed{D}^{KW} \quad\mathrm{chiral~symmetry,}\\
&D^{KW\dag}D^{KW} = (m - \slashed{D}^{KW}) (m + \slashed{D}^{KW})
= m^2-\slashed{D}^{{KW}^2} = D^{KW} D^{KW\dag} \quad\mathrm{normality~and}\\
&D^{KW\dag} D^{KW}\gamma_5 =
D^{KW\dag} \gamma_5 D^{KW\dag}= 
\gamma_5 D^{KW} D^{KW\dag}= 
\gamma_5 D^{KW\dag} D^{KW}. \label{LR}
\end{align}
The consequences are that the eigenvalue spectrum of $\slashed{D}^{KW}$ falls on the imaginary axis with pairs of complex conjugate eigenvalues $\pm i\lambda$,
the operator $D^{KW\dag} D^{KW}$ is positive definite with eigenvalues $m^2+\lambda^2$ and Eq. (\ref{LR}) means that the eigenvalues of $D^{KW\dag} D^{KW}$ are either left(L) - or right(R)-handed. Thus, we can write $D^{KW}$ as
\begin{equation}
D^{KW} = \left(
\begin{array}{cc}
D^{KW}_{LL}&D^{KW}_{LR}\\
D^{KW}_{RL}&D^{KW}_{RR}
\end{array}
\right)
=\left(
\begin{array}{ccc|ccc}
&m\mathbf{1}&&&\slashed{D}_{LR}^{KW}&\\
\hline
&\slashed{D}_{RL}^{KW}&&&m\mathbf{1}&\\
\end{array}
\right)
\end{equation}
and $D^{KW\dag} D^{KW}$ will have a block diagonal form
\begin{equation}
D^{KW\dag} D^{KW}= 
\left(
\begin{array}{cc}
m^2 - \slashed{D}^{KW}_{LR}\slashed{D}^{KW}_{RL} & 0\\
0&m^2 - \slashed{D}^{KW}_{RL}\slashed{D}^{KW}_{LR} 
\end{array}
\right),
\end{equation}
where the determinants of two blocks have the property $\det(D^{KW\dag} D^{KW})_{LL} = \det (D^{KW\dag} D^{KW})_{RR}$. Then the pseudofermion action in the HMC algorithm will have the following form
\begin{equation}
S_f=\chi_L^\dag[(D^{KW\dag} D^{KW})_{LL}]^{-N_f/2}\chi_L
\end{equation}
where $\chi_L$ means the left handed components of the pseudofermions and $N_f$ is the number of flavors. We have two flavors $(N_f=2)$ with the KW action, thus, we do not need to perform rooting. We can simply use a conjugate gradient (CG) solver for two Dirac components (with the half vector trick this reduces to one) when calculating the fermion force 
$
\partial\chi^+_L (D^{KW\dag} D^{KW})_{LL}^{-1}(U) \chi_L/ \partial [U_\mu(x)].
$
In the next Section we show the results of our tuning procedure of the counter-term coefficients.
\section{The non-perturbative tuning of the counter-term coefficients} \label{tune}
\begin{figure}
 \centering 
\includegraphics[width=.495\textwidth]{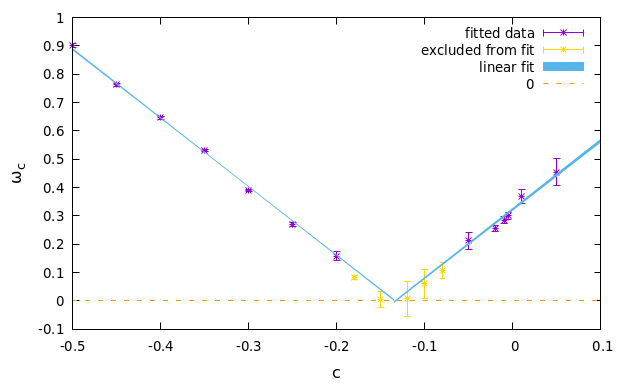} \includegraphics[width=.495\textwidth]{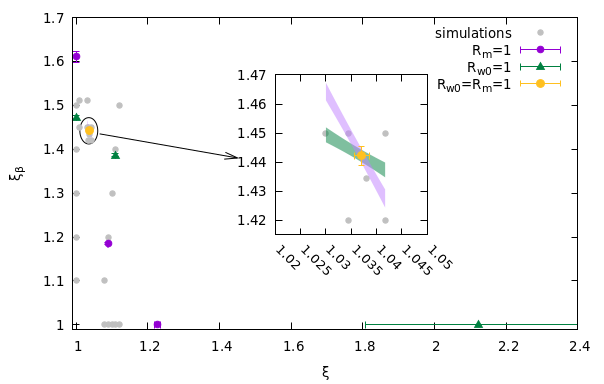}
           \caption{The reduced frequency of the correlator in the $\gamma_0$ channel as a function of the parameter $c$ (left hand side). The yellow points are excluded from the linear fit (blue band). Two dimensional scan of the $\xi_\beta- \xi$ plane (right hand side). Grey points show the parameter values of different simulations and the yellow point shows the result of the tuning. Purple and green points and bands show where the criteria A1 and A2 is satisfied respectively. \label{fig:Vshape} }
\end{figure}

\begin{figure}
 \centering 
\includegraphics[width=.495\textwidth]{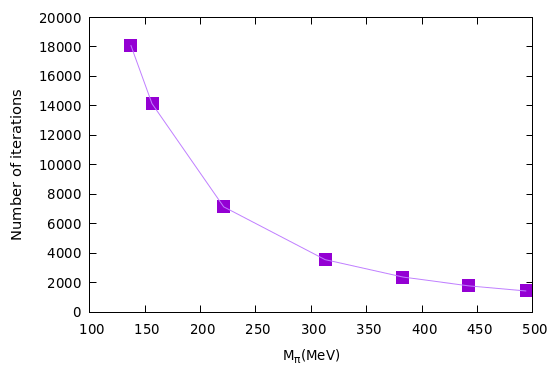} \includegraphics[width=.495\textwidth]{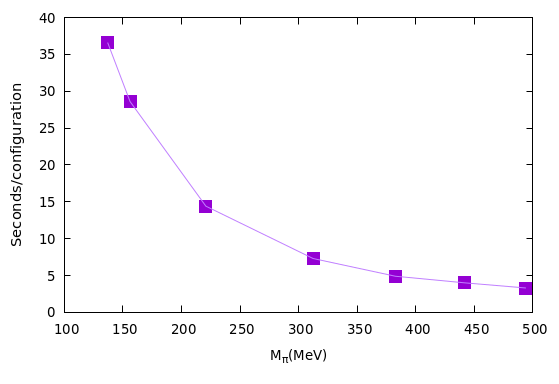}
           \caption{Number of iterations needed for the MCG solver with precision $10^{-8}$ as a function of the physical pion mass (left hand side). Seconds needed on a Juwels-Booster node with 4$\times$A100 Gpus to generate a configuration as a function of the physical pion mass (right hand side). \label{fig:mass_scan} }
\end{figure}

We applied Hasenbusch-preconditioning \cite{Hasenbusch_2001} and for the preconditioned field we used a Rational HMC (RHMC) \cite{Clark_2007} with a Multi-shift \cite{MSsolver} CG (MCG)  solver.
 We tuned the parameters $c$, $\xi\equiv (1+d)$ and $\xi_\beta\equiv (1+d_G)$ non-perturbatively to restore the explicitly broken anisotropy of the lattice.
The parameters of the simulations in lattice units were $\beta=3.5$, temporal extension $N_t=64$, spatial volume $V=16^3$. The spatial lattice spacing in physical units was $a=0.180(3)~\mathrm{fm}$ and we used approximately $200$ lattice configurations for each setup of counter-term coefficients.
We first tuned the parameter $c$ using the mesonic correlation function in the $\gamma_0$ channel
 \begin{equation} \label{C0}
    C_{i}(x,y) \sim \langle \bar\psi(x)\gamma_i\psi(x) \bar\psi(y)\gamma_i\psi(y)\rangle \qquad (i=0),
\end{equation} following the method that was demonstrated in quenched QCD in Ref.~\cite{weber2016correlation}. When $c$ is not tuned correctly the correlator has an oscillatory behavior in the direction of the anisotropy
\begin{equation} \label{C0t}
C_0(t)\approx
A\left[
\cos(t\omega+\phi) \exp(-mt)
+
\cos((N_t-t)\omega+\phi) \exp(-m(N_t-t))
\right]
\end{equation}
where $\omega$ and $\phi$ are the frequency and phase of the oscillatory term and $m$ is the mass corresponding to the ground state of the correlator. Changing the parameter $c$ shifts the frequency and the oscillation disappears when $\omega=\pi$. We did a scan in $c$ and at each value we measured the correlator in Eq. (\ref{C0}) in time direction and fitted the function Eq. (\ref{C0t}) to extract the frequency $\omega$. We interpolated the value of the reduced frequency $\omega_c\equiv \omega-\pi$ as a function of $c$ to find its zero. This is demonstrated in the left hand side plot of Fig. \ref{fig:Vshape}.
We performed a two dimensional scan To find the tuned values of $\xi$ and $\xi_\beta$, while keeping the previously tuned value of $c$ fixed as it did not depend significantly on the other two parameters. For the simultaneous tuning of the other two parameters we checked the following two criteria
\begin{itemize}
\item[A1] The pseudo-Goldstone propagator in the $\gamma_5$ channel ($i=5$ in Eq. (\ref{C0}) measured in parallel and perpendicular directions to time has to give the same mass $R_m=m_{||}/m_{\perp}=1.$
\item[A2] The lattice spacing in the directions parallel and perpendicular to time have to be the same $R_a=a_{||}/a_{\perp}=1.$
\end{itemize}
We used the $w_0$ scale \cite{borsanyi2012} to find the physical values of the lattice spacings. We did a scan along the $\xi$  and the $\xi_\beta$ axis and in the diagonal direction by pairing the coordinates in ascending order. We calculated $R_m$ and $R_a$ and interpolated the coordinates $\xi$ and $\xi_\beta$ where they would reach unity for each of the three scan directions. We then could roughly estimate the intersection of the two curves that satisfy criteria A1 and A2. We zoomed in the vicinity of the estimated intersection and performed several simulations in that region. By fitting planes on the values of $R_a$ and of $R_m$ we could determine the intersection where both A1 and A2 are satisfied. This is demonstrated in Fig \ref{fig:Vshape}. Our results for the tuned values at $\beta=3.5$ are $c=-0.1336(13),~ \xi=1.0370(15),~ \xi_\beta=1.442(3)$. We also did a mass scan with the setting $N_t=48, V=24^3$ to check the performance of the code toward the physical point. This was done simultaneously with the tuning procedure, thus the setup $\xi=1.0381,~ \xi_\beta=1.4345$ was used with $c=-0.135$. In this case the deviation of $R_a$ and $R_m$ from unity is $0.1\%$. The results are shown in Fig. \ref{fig:mass_scan}.
We found that doing simulations at physical point would be still manageable using minimally doubled fermions with tuned parameters.
\section{Conclusions and outlook} \label{summary}
We implemented a new dynamical code using a specific type of minimally doubled fermion discretization the Naik improved Karsten-Wilczek fermions. We introduced a tree-level improvement of the fermion action with three-hop Naik terms. Minimally doubled fermions are anisotropic hence there have to be additional counter-terms to restore isotropy.  We performed the first dynamical simulations with minimally doubled fermions and tuned the coefficients of the counter-terms non-perturbatively. We tested the performance of the code with a parameter setting close to the tuned values whit decreasing bare fermion masses. The performance remained reasonable for dynamical simulations even at the physical pion mass setting. Thus the Karsten-Wilczek action is a promising discretization for dynamical simulations.

It is always desirable to reduce the effects of finite lattice spacing. A way to improve he action further is to determine higher order terms by using perturbation theory to realize an improvement beyond tree level. A well improved action with fine tuned parameters then can be used for thermodynamical lattice simulations with two degenerate flavors. 
\section*{Acknowledgments}
R. Vig was funded by the DFG under the Project No. 496127839.
This work is also supported by the MKW NRW under the funding code NW21-024-A.
The authors gratefully acknowledge the Gauss Centre for
Supercomputing e.V. (\url{www.gauss-centre.eu}) for funding
this project by providing computing time on the GCS
Supercomputers Juwels-Booster at Juelich Supercomputer
Centre.

\end{document}